# Practical deviational particle method for variance reduction in polyatomic gas DSMC simulations

Takehiro Shiraishi and Ikuya Kinefuchi

## Abstract

The direct simulation Monte Carlo (DSMC) method is a widely used stochastic particle approach to solving the Boltzmann equation. However, its computational cost remains a major drawback, which can be attributed to statistical errors when handling flows with low Mach numbers. Thus, many studies have focused on variance reduction to reduce the computational cost. One approach is the deviational particle (DP) method, which focuses solely on modeling deviations from the equilibrium state. The DP method has been implemented in the low-variance deviational simulation Monte Carlo (LVDSMC) method, which has proven effective for monatomic gas simulations but faces limitations when extended to polyatomic gases. In this study, we present a practical DP method for polyatomic gas simulations that combines the LVDSMC method with the Larsen–Borgnakke (LB) model, which introduces a group reduction algorithm for the inelastic collision process. Numerical experiments demonstrated that the proposed method efficiently and accurately simulates flows across a relatively broad range of non-equilibrium values. Remarkably, the variance was reduced to about 5% that of the DSMC method.

## Introduction

Advances in microelectromechanical systems (MEMS) and nanoelectromechanical systems (NEMS) have led to growing interest in nano-/microfluidic flows with low Mach numbers. The Navier–Stokes equation is inadequate for describing these flows owing to their rarefied nature, and they are instead governed by the Boltzmann equation. Various methods for solving the Boltzmann equation have been proposed, which can be classified into two categories: analytical and numerical methods. Analytical methods introduce linearization or approximations to simplify the problem. Such methods include approximate kinetic models such as the Fokker–Planck approximation to the Boltzmann collision operator [1], the linearized Boltzmann equation method [2,3], the moment method [4,5], and the Bhatnagar–Gross–Krook (BGK) equation method [6]. Analytical methods are useful when the approximations are valid in the flows of interest, but often these approximations are so crude that they lack generality.

Numerical methods estimate the solution through iterative processes and gradually refine the results by successively improving the approximations, which is typically by using discrete steps in time and space. The direct simulation Monte Carlo (DSMC) method [7] is the most widely used because of its simplicity and stability. However, statistical fluctuations are a well-known drawback, particularly for flows with a low Mach number. The statistical fluctuations inherent to the DSMC method decrease with the inverse square root of the sample size, so an extremely large sample size is required to obtain smooth results, but this leads to a prohibitive computational cost. Several approaches have been proposed to improve the computational efficiency for these flows, including discrete velocity [8,9], finite difference [10,11], and variance reduction methods [12–17]. Discrete velocity or discrete ordinate methods significantly simplify calculations by assuming that the velocity space has only a finite number of discrete values. Finite element methods use the Monte Carlo method of quadrature to calculate the collision integral and solve the Boltzmann equation with a finite difference method of computational fluid dynamics. However, these two methods introduce some simplifications that may reduce accuracy.

Variance reduction methods are a newer approach that reduces statistical fluctuations while maintaining the advantages of the DSMC method. Some variance reduction methods utilize importance weights to exploit the correlation between equilibrium and non-equilibrium simulations [16,17]. The deviational particle (DP) method reduces statistical errors by simulating only deviations from the equilibrium by using sample particles with signs. Although the DP method can significantly reduce statistical fluctuations, it tends to generate redundant particles (i.e., pairs of sample particles with similar physical properties but different signs) that have little impact on the estimation of distribution functions [15]. Thus, particle cancelation in the collision process is required for stable computation, which makes true practicality difficult to achieve. The low-variance deviational simulation Monte Carlo (LVDSMC) method [12,13,15], has overcome the adversity by utilizing the Hilbert form of the collision operator, which results in a source–sink formulation of the collision process that estimates the probability of generating redundant particles in a collision event and rejecting the event at this rate. By incorporating pre-canceling, the LVDSMC method circumvents the need for additional cancelation processes, which allows it to efficiently simulate low-signal flows of monatomic gases. However, a major limitation of the LVDSMC method is that it remains inapplicable to polyatomic gas flows.

Here, we propose a new DP method specifically designed for polyatomic gas flows. The proposed method incorporates a model that calculates collision terms for polyatomic molecules called the Larsen–Borgnakke (LB) model and a group reduction algorithm to maintain a reasonable computational cost. To simulate a polyatomic gas flow, the proposed method only considers the rotational energy and translational energy. Because the proposed method is intended for microfluidic flows, these molecules do not fall within the temperature range where vibrational energy is excited. A hard sphere model is employed to calculate molecular interactions. In this paper, we present the formulation of the proposed method and its application in a numerical experiment where we compared its accuracy and efficiency to that of the DSMC method.

# Deviational particle method for polyatomic gas flows

The governing equation for the DP method is derived directly from the Boltzmann equation [15].

$$\frac{\partial f}{\partial t} + \mathbf{c}\frac{\partial f}{\partial \mathbf{x}} = \mathcal{Q}[f,f], \tag{1}$$

where $f$ is a distribution function of the velocity or rotational energy, $t$ is the time, $\mathbf{c}$ is the velocity, and $\mathbf{x}$ is the spatial coordinate. $\mathcal{Q}[f,f]$ is the collision term representing the effects of molecular interactions on $f$. In the DP method, the distribution function is decomposed into the reference distribution $f_0$ and deviational distribution $f_d$:

$$f = f_0 + f_d. \tag{2}$$

The reference distribution $f_0$ is a prescribed parameter that can be defined arbitrarily. In this paper, we use equilibrium distributions to simplify the governing equation. The reference distribution function of the velocity is the Maxwell–Boltzmann distribution, and the reference distribution function of the rotational energy is the Boltzmann distribution:

$$f_0(\mathbf{c})d\mathbf{c} = \frac{n_0}{\pi^{\frac{3}{2}} c_0^3} \exp\left(-\frac{\|\mathbf{c}-\mathbf{u}_0\|^2}{c_0^2}\right) d\mathbf{c}, \tag{3}$$

$$f_0(\varepsilon)d\varepsilon = \frac{\varepsilon^{\frac{\xi}{2}-1}}{(k_b T_0)^{\frac{\xi}{2}} \cdot \Gamma\left(\frac{\xi}{2}\right)} \exp\left(-\frac{\varepsilon}{k_b T_0}\right) d\varepsilon \tag{4}$$

where $n_0$ is the number density, $\mathbf{u}_0$ is the mean velocity, $c_0 = \sqrt{2RT_0}$ is the most probable velocity, and $T_0$ is the temperature. All of these parameters are in the equilibrium state. Substituting the definition of the deviational distribution in equation (2) into the Boltzmann equation in equation (1) yields the governing equation for the DP method:

$$\frac{\partial f_d}{\partial t} + \mathbf{c}\frac{\partial f_d}{\partial \mathbf{x}} = \mathcal{L}[f_d, f_0] + \mathcal{Q}[f_d, f_d]. \tag{4}$$

The collision operator on the right side has a linear term $\mathcal{L}[f_d, f_0]$ and nonlinear term $\mathcal{Q}[f_d, f_d]$. $\mathcal{L}[f_d, f_0]$ and $\mathcal{Q}[f_d, f_d]$ correspond to changes in the deviational distribution $f_d$ due to interactions between deviational particles and the reference distribution and collisions between two deviational particles, respectively. Similar to the DSMC method, the DP method calculates the time evolution of the deviational velocity distribution by splitting the governing equations into collision and advection steps. The implementation of the processes necessary for polyatomic gas simulation is described below.

## Particle generation

The deviational distribution can take negative values depending on the reference distribution, so it is represented by particles with plus or minus signs. We extended the generation algorithms presented by Wagner [15] to consider rotational energy. The generation from $f_d$ is performed by using the acceptance–rejection method [18,19] with the envelope function $f + f_0$:

1. Sample velocity $\mathbf{c}$ from the probability distribution $f(\mathbf{c}) + f_0(\mathbf{c})$.
2. Sample rotational energy $\varepsilon$ from the probability distribution $f(\varepsilon) + f_0(\varepsilon)$.
3. Generate a particle with velocity $\mathbf{c}$ and rotational energy $\varepsilon$ at the following probability:

$$P(\mathbf{c}, \varepsilon) = \frac{|f(\mathbf{c})f(\varepsilon) - f_0(\mathbf{c})f_0(\varepsilon)|}{f(\mathbf{c})f(\varepsilon) + f_0(\mathbf{c})f_0(\varepsilon)}. \tag{6}$$

   The sign of this particle is same as the sign of $f(\mathbf{c})f(\varepsilon) - f_0(\mathbf{c})f_0(\varepsilon)$.

4. Iterate steps 1–3 for $n/g$ times, where $n$ is the number density of the initial condition and $g$ is the particle weight.

Sampling from $f + f_0$ in steps 1 and 2 is implemented in a two-step process. First, $f$ is sampled with the probability $P_f = \int_{R^3} f d\mathbf{c} / (\int_{R^3} f d\mathbf{c} + \int_{R^3} f_0 d\mathbf{c})$. Second, $f_0$ is sampled with the probability $1 - P_f$. The Box–Muller method [20] is used to sample from the Maxwell–Boltzmann distributions.

## Advection step

The advection step involves calculating the left-hand side of the governing equation (6):

$$\frac{\partial f_d}{\partial t} + \mathbf{c}\frac{\partial f_d}{\partial \mathbf{x}} = 0, \tag{7}$$

The above equation (7) shows that deviational particles are advected in the same manner as in the original DSMC method. Therefore, the advection of a deviational particle $i$ in a discretized timestep $\Delta t$ can be simply implemented as follows:

$$\mathbf{x}_i(t + \Delta t) = \mathbf{x}_i(t) + \mathbf{c}_i(t)\Delta t. \tag{8}$$

## Boundary condition

The general boundary condition of the DSMC method is given by [7]

$$(\mathbf{c} \cdot \mathbf{n})f(\mathbf{c}) = -\int_{R_{\text{out}}} (\mathbf{c}' \cdot \mathbf{n})f(\mathbf{c}')R(\mathbf{c}; \mathbf{c}')d\mathbf{c}' + (\mathbf{c} \cdot \mathbf{n})f_{\text{in}}(\mathbf{c}). \tag{8}$$

where $R(\mathbf{c}; \mathbf{c}')$ is the scattering kernel relating the velocity of the impinging particle $\mathbf{c}'$ and the velocity of the reflected particle $\mathbf{c}$, $\mathbf{n}$ is a unit normal vector indicating the direction from the boundary into the computational region, and $R_{\text{out}}$ is the velocity space that the outflow can take at the boundary surface. The first term on the right side represents the reflection at the boundary, and the second term represents the influx from the boundary. To simulate deviational particles, this DSMC boundary condition is modified with equation (2) to obtain:

$$(\mathbf{c} \cdot \mathbf{n}) f_d(\mathbf{c}, \varepsilon) = \int_{R_{\text{out}}} \int_0^\infty (\mathbf{c}' \cdot \mathbf{n}) f_d(\mathbf{c}', \varepsilon') R(\mathbf{c}, \varepsilon; \mathbf{c}', \varepsilon') d\mathbf{c}' d\varepsilon' + (\mathbf{c} \cdot \mathbf{n}) f_{\text{in}}(\mathbf{c}, \varepsilon)$$
$$+ \int_{R_{\text{out}}} \int_0^\infty (\mathbf{c}' \cdot \mathbf{n}) f_0(\mathbf{c}', \varepsilon') R(\mathbf{c}, \varepsilon; \mathbf{c}', \varepsilon') d\mathbf{c}' d\varepsilon' - (\mathbf{c} \cdot \mathbf{n}) f_0(\mathbf{c}, \varepsilon), \quad (10)$$

More specific boundary conditions are derived in our previous paper [21].

## Collision process

For the collision process, the time-evolution term and collision terms of equation (6) are simulated:

$$\frac{\partial f_d}{\partial t} = \mathcal{L}[f_d, f_0] + \mathcal{Q}[f_d, f_d]. \quad (11)$$

Here, we assume that $f_d \ll f_0$, so we only consider the linear part $\mathcal{L}[f_d, f_0]$. This is a reasonable assumption because our interest is in nano-/microfluidic flows with low Mach numbers, where the changes in the distribution $f$ are relatively small. Consequently, equation (10) contracted to (11).

$$\frac{\partial f_d}{\partial t} = \mathcal{L}[f_d, f_0]. \quad (12)$$

Therefore, we only calculate collisions between one particle representing the reference distribution (i.e., reference particle) and another particle representing the deviational distribution (i.e., deviational particle). The general method for calculating a collision between a deviational particle $i$ and reference particle $j$ is described below as per Wagner [15]:

1. Choose the colliding deviational particle $i$, which has a spatial coordinate $\mathbf{x}_i$, velocity $\mathbf{c}_i$, rotational energy $\varepsilon_i$, and sign $\sigma_i$.

2. Sample the velocity $\mathbf{c}_j$ and rotational energy $\varepsilon_j$ from the reference distribution as the properties of the reference particle $j$.

3. Calculate the post-collision values $\mathbf{c}'_i, \mathbf{c}'_j, \varepsilon'_i, \varepsilon'_j$ with a certain collision operator.

4. To reflect the changes in the probability distributions of the velocity and internal energy, generate particles $(\mathbf{x}_i, \mathbf{c}'_i, \varepsilon'_i, \sigma_i), (\mathbf{x}_j, \mathbf{c}'_j, \varepsilon'_j, \sigma_i), (\mathbf{x}_i, \mathbf{c}_j, \varepsilon_j, -\sigma_i)$, and remove particle $i$.

Note that an additional particle with an opposite sign is generated when a reference particle collides. Therefore, particles increase by two per collision, which leads to an eventual explosion of the number particles. This issue can be solved by introducing a cancelation technique either before or after the collision. For practical implementation, two critical considerations emerge: the calculation of post-collision values for polyatomic collisions and the cancelation of redundant particles. The proposed method addresses the former by integrating the LB model [22] as a collision operator within the framework of the DP method outlined earlier. The LB model characterizes polyatomic collisions as a blend of elastic and inelastic collisions. The ratio of inelastic collisions over all collisions $\phi$ is a model parameter that is prescribed as a material property. Elastic collisions involve no energy exchange between different modes while inelastic collisions incorporate energy exchange via a specific algorithm. The proportion of elastic collisions serves as the sole parameter determined empirically to establish the relaxation rate. In the LB model, elastic collisions are calculated as monatomic molecular collisions. Inelastic collisions between particle $i$ and particle $j$ are calculated according to the following algorithm [22]:

1. Calculate the sum of the relative translational energy $E_t$ and rotational energy $E_r$:

$$E_c = E_t + E_r = \frac{1}{2} m_r \|\mathbf{c}_i - \mathbf{c}_j\|^2 + \varepsilon_i + \varepsilon_j, \tag{13}$$

   where $m_r = m/2$ is the reduced mass.

2. Sample the post-collision relative translational energy $E_t'$ and rotational energy $E_r'$ from the following probability distribution:

$$f\left(\frac{E_t'}{E_c}\right) \propto \left(\frac{E_t'}{E_c}\right)^{\frac{3}{2}-\omega} \left(1 - \frac{E_t'}{E_c}\right)^{\xi-1}, \tag{14}$$

   where $E_r' = E_c - E_t'$ and $\omega$ is the temperature power exponent of the viscosity of the gas. Here, the distribution functions of velocity and rotational energy are assumed as equilibrium distributions with the effective temperature $T = \frac{E_c}{k\left(\frac{5}{2}-\omega\right)}$.

3. Sample the post-collision rotational energies of particle $i$ and particle $j$ from the following probability distribution:

$$f\left(\frac{\varepsilon_i'}{E_i'}\right) \propto \left(\frac{\varepsilon_i'}{E_i'}\right)^{\frac{\xi}{2}-1} \left(1 - \frac{\varepsilon_i'}{E_i'}\right)^{\frac{\xi}{2}-1}, \tag{15}$$

   where $\varepsilon_j' = E_i' - \varepsilon_i'$.

4. Calculate the post-collision velocities by using the obtained post-collision relative translational energy $E_t'$:

$$c'_i = \frac{1}{2}\left(c_i + c_j + \sqrt{\frac{2E'_t}{m_r}}\, e\right), \qquad (16)$$

$$c'_j = \frac{1}{2}\left(c_i + c_j - \sqrt{\frac{2E'_t}{m_r}}\, e\right), \qquad (17)$$

where **e** is a random unit vector.

Although this artificial relaxation process of the LB model is nonphysical in the context of state-specific relaxation, it is exceedingly good at reproducing real relaxation behavior, such as the temperature behind a shock wave [23,24]. To address the latter consideration, we employ distinct techniques for elastic and inelastic collisions. The source–sink collision process from the LVDSMC method is applied to elastic collisions in the LB model because they do not exchange energy between different energy modes and are equivalent to monatomic collisions. For inelastic collisions, we introduce an explicit cancelation technique that we call the group reduction algorithm immediately after a collision is calculated with the general collision process described above. This strategy is straightforward and helps prevent system blow-up for complicated inelastic collisions involving energy exchanges. Figure 1 summarizes the collision process in the proposed method.

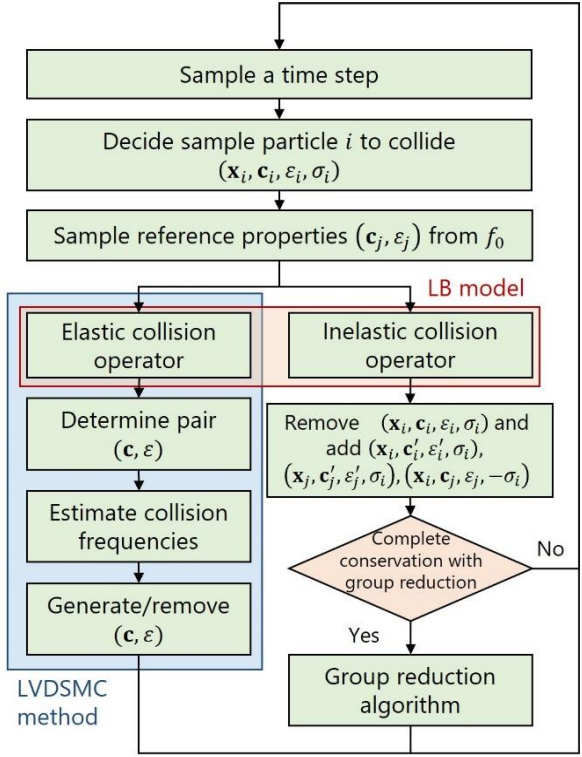

**Fig. 1** Flowchart of the collision process.

The detailed implementation is described below.

1. Advance time based on an exponential distribution with the following parameter:

$$\sum_{i=1}^{N_{\text{cell}}} \int_{R^3} (4 - 3\phi)\pi d^2 |c_i - w| f_0(w) dw, \tag{18}$$

where $N_{\text{cell}}$ is the number of sample particles in the computational cell. The collision step is terminated when the time step is reached.

2. Choose a colliding particle $i$ with the probability

$$\frac{\int_{R^3} \pi d^2 |c_i - w| f_0(w) dw}{\sum_{m=1}^{N_{\text{cell}}} \int_{R^3} \pi d^2 |c_m - w| f_0(w) dw}. \tag{19}$$

3. Determine the position of the reference particle $j$ as a random position in the computational cell, and determine the velocity and rotational energy according to the following probability density distributions:

$$f(c_j) = \frac{|c_i - c_j| f_0(c_j)}{\int_{R^3} du \, |c_i - u| f_0(u)}, \tag{20}$$

$$f(\varepsilon_j) = \frac{f_0(\varepsilon_j)}{\int_0^\infty d\varepsilon f_0(\varepsilon)}. \tag{21}$$

4. Process the collision as an elastic collision with the probability $4(1 - \phi)/(4 - 3\phi)$ and as an inelastic collision with the probability $\phi/(4 - 3\phi)$.

5. Return to Step 1.

The collision operators in the above collision process are described below.

## Elastic collision operator

The source–sink process prevents the generation of redundant particles by statistically estimating the probability of this generation and rejecting particle generation at this probability [15]. To enable the analytical estimation of the probability, the source–sink process considers only one velocity at a time out of the four velocities involved in a collision; as a complement, it considers collision events at four times the actual collision number. We extended this source–sink process to calculate the elastic collisions of polyatomic gas molecules by estimating the probability of generating a specific pair of velocity and rotational energy $(c, \varepsilon)$ rather than just the velocity. The collision algorithm for the deviational particle $i$ is implemented as follows:

1. Calculate the post-collision variables $c_i', c_j', \varepsilon_i', \varepsilon_j'$ by using the hard sphere collision rule:

$$c_i' = \frac{1}{2}(c_i + c_j + e|c_i - c_j|), \tag{22}$$

$$c_j' = \frac{1}{2}(c_i + c_j - e|c_i - c_j|) \tag{23}$$

$$\varepsilon'_i = \varepsilon_i, \qquad \varepsilon'_j = \varepsilon_j, \tag{24}$$

where **e** is a random unit vector.

2. Generate a random integer number $k$ from 1 to 4 with a uniform probability. Go to Step 5 if $k = 4$ and Step 3 otherwise. Choose a pair of velocity and rotational energy to generate $(\mathbf{c}, \varepsilon)$ as $(\mathbf{c}'_i, \varepsilon'_i)$ if $k = 1$, $(\mathbf{c}'_j, \varepsilon'_j)$ if $k = 2$, and $(\mathbf{c}_j, \varepsilon_j)$ if $k = 3$.

3. Calculate the probability of generating a particle with the properties $(\mathbf{c}, \varepsilon)$, $P(\mathbf{c}, \varepsilon) = |S|/S_{\max}$. $S$ is the net increase per unit time in particles with the variables $(\mathbf{c}, \varepsilon)$, and $S_{\max}$ is the expected number of collisions per unit time involving particles with these variables:

$$S = \sum_{m \neq i}^{N} \sigma_m [\alpha_1(m, \varepsilon) K_1(m, \mathbf{c}) + \alpha_2(\varepsilon) K_2(m, \mathbf{c}) + \alpha_3(\varepsilon) K_3(m, \mathbf{c})], \tag{25}$$

$$S_{\max} = \sum_{m \neq i}^{N} \sigma_m [\alpha_1(m, \varepsilon) K_1(m, \mathbf{c}) + \alpha_2(\varepsilon) K_2(m, \mathbf{c}) + |\alpha_3(\varepsilon) K_3(m, \mathbf{c})|]. \tag{26}$$

$K_1(m, \mathbf{c})$ is the frequency that the particle $m$ collides and has the post-collision velocity $\mathbf{c}'_m = \mathbf{c}$, $K_2(m, \mathbf{c})$ is the frequency that the rotational energy sampled from $f_0$ is involved in a collision and has the post-collision velocity $\mathbf{c}$, and $K_3(m, \mathbf{c})$ is the frequency that the velocity sampled from $f_0$ is $\mathbf{c}$. These functions are from the source–sink process in the LVDSMC method [15]:

$$K_1(m, \mathbf{c}) = K_2(m, \mathbf{c}) = \frac{1}{|\mathbf{c}_m - \mathbf{c}|} \int_{\Gamma(\mathbf{c}_m - \mathbf{c})} d\mathbf{u} f_0(\mathbf{c} + \mathbf{u}), \tag{27}$$

$$K_3(m, \mathbf{c}) = -\pi f_0(\mathbf{c}) |\mathbf{c}_m - \mathbf{c}|, \tag{28}$$

where $\Gamma(\mathbf{c})$ is a plane through the origin perpendicular to $\mathbf{c}$.

This method considers the rotational energy by introducing $\alpha$ functions that calculate the probabilities of generating or removing particles with the rotational energy $\varepsilon$. $\alpha_1(m, \varepsilon) d\varepsilon$ represents the probability that the particle $m$ collides and has the post-collision rotational energy $\varepsilon'_m = \varepsilon$, $\alpha_2(m, \varepsilon) d\varepsilon$ is the probability that the rotational energy sampled from $f_0$ is involved in a collision and has the post-collision rotational energy $\varepsilon$, and $\alpha_3(\varepsilon)$ is the frequency that the rotational energy sampled from $f_0$ is $\varepsilon$:

$$\alpha_1(m, \varepsilon) = \delta(\varepsilon_m - \varepsilon'), \tag{29}$$

$$\alpha_2(\varepsilon) = \alpha_3(\varepsilon) = f_0(\varepsilon'), \tag{30}$$

For implementation, the delta function in equation (29) is approximated by a rectangular function with the width $\delta\varepsilon$:

$$\alpha_1(m, \varepsilon) \approx H\left(\varepsilon_m - \varepsilon + \frac{\delta\varepsilon}{2}\right) - H\left(\varepsilon_m - \varepsilon - \frac{\delta\varepsilon}{2}\right), \tag{31}$$

where $H$ is the Heaviside step function. $\delta\varepsilon$ is a simulation constant that is prescribed by the user. Because elastic collisions do not exchange energy between translational and rotational modes, the collision frequencies to generate/remove particles with the properties $(\mathbf{c}, \varepsilon)$ can be obtained simply by multiplying the $K$ functions and $\alpha$ functions. Consequently, the first two terms in equation (25) represent the expected increase in particles with variables $(\mathbf{c}, \varepsilon)$ generated as post-collision particles per unit time, and the third term is the expected decrease in particles with variables $(\mathbf{c}, \varepsilon)$ removed after collisions per unit time.

4. Generate the particle $(\mathbf{x}, \mathbf{c}, \varepsilon, \text{sign}(S))$ with the probability $P(\mathbf{c}, \varepsilon)$. The position of the particle $\mathbf{x}$ is randomly sampled from the computational cell.

5. Remove particle $i$.

## Inelastic collision operator

The post-collision properties in the inelastic collision process are calculated by the LB model as the redistribution of energy. Because explicitly determining the probability of generating or removing particles with specific properties during a collision is nearly impossible owing to the complexity arising from energy exchange between different energy modes, we instead employ a straightforward collision implementation. First, we generate three particles $(\mathbf{x}_i, \mathbf{c}'_i, \varepsilon'_i, \sigma_i)$, $(\mathbf{x}_j, \mathbf{c}'_j, \varepsilon'_j, \sigma_i)$, and $(\mathbf{x}_i, \mathbf{c}_j, \varepsilon_j, -\sigma_i)$ as described above. Then, we introduce a new cancelation technique that we call the group reduction algorithm to control the number of sample particles. The central concept behind this algorithm is to cancel out the particle $(\mathbf{x}_i, \mathbf{c}_j, \varepsilon_j, -\sigma_i)$ additionally generated by the collision process by pairing it with a sample particle existing in the same computational cell and possessing similar properties with a different sign $(\mathbf{x}_k, \mathbf{c}_k, \varepsilon_k, \sigma_i)$. The similarity of the two particles is defined as the distance in velocity space $|\mathbf{c}_j - \mathbf{c}_k|$. The rotational energy is not considered because finding a similar particle becomes much more difficult with its inclusion. Because the rotational energy $\varepsilon_k$ of this additional particle is not considered in the similarity calculation, it should not be involved in the cancelation. Thus, one of the remaining two particles has the rotational energy $\varepsilon_k$, and the other particle has the rotational energy $\varepsilon'_i + \varepsilon'_j - \varepsilon_j$ to satisfy the conservation of rotational energy. Although this cancelation process looks harsh and harms the simulation accuracy, it can be regarded as an approximation of the redistribution of energy by the LB model. Namely, while the original model uses a two-step distribution of energy, where the total energy is distributed to translational and rotational energy modes and the energy of each mode is then distributed to two sample particles, the group reduction algorithm only conducts the first distribution and adopts cancelation to conserve the results of the distribution. For velocity, a simple removal of particle $j$ and the selected particle $k$ significantly violates conservation laws because of the limited number of sample particles in a cell and the potential dissimilarity of the two particles. The velocities of the remaining particles $(\mathbf{x}_i, \mathbf{c}'_i, \varepsilon'_i, \sigma_i)$ and $(\mathbf{x}_j, \mathbf{c}'_j, \varepsilon'_j, \sigma_i)$ are therefore updated subsequently to ensure the conservation of translational energy in each direction. Sometimes, conservation is impossible to obtain by the group reduction algorithm. In the case, reduction is not performed, and the three particles are added to the system. The algorithm is presented below:

1. Calculate the post-collision variables with the LB model. Three particles $(\mathbf{x}_i, \mathbf{c}'_i, \varepsilon'_i, \sigma_i)$, $(\mathbf{x}_j, \mathbf{c}'_j, \varepsilon'_j, \sigma_i)$, and $(\mathbf{x}_i, \mathbf{c}_j, \varepsilon_j, -\sigma_i)$ are generated.

2. Choose particle $k$ from the same computational cell with the same sign as particle $i$ and such that $|\mathbf{c}_j - \mathbf{c}_k|$ is the smallest. Then, the three particles generated in Step 1 and this particle $k$ make up a group.

3. From the group defined in Step 2, generate two representative particles with the same sign as particle $i$: particle $w$ and $l$. The velocities and rotational energies of these particles are calculated as follows.

$$\mathbf{c}_w = \mathbf{V} + \sqrt{3T_g}\mathbf{e}, \qquad \varepsilon_w = \varepsilon'_i + \varepsilon'_j - \varepsilon_j, \tag{32}$$

$$\mathbf{c}_l = \mathbf{V} - \sqrt{3T_g}\mathbf{e}, \qquad \varepsilon_l = \varepsilon_k, \tag{33}$$

where $\mathbf{V}, T_g, \mathbf{e}$ are given as

$$\mathbf{V} = \frac{\mathbf{c}'_i + \mathbf{c}'_j - \mathbf{c}_j + \mathbf{c}_k}{2}, \tag{34}$$

$$T_g = \frac{1}{3}\left(\frac{{c'_i}^2 + {c'_j}^2 - c_j^2 + c_k^2}{2} - V^2\right), \tag{35}$$

$$e_n = \sqrt{\frac{T_n}{3T_g}}, \qquad T_n = \frac{{c'_{i,n}}^2 + {c'_{j,n}}^2 - c_{j,n}^2 + c_{k,n}^2}{2} - V_n^2, \tag{36}$$

for $n = 1,2,3$. Negative $T_n$ for any $n$ means complete conservation is not possible. In this case, three particles are generated directly from step 1.

Note that the results can converge to the DSMC simulation results, which are known to be equivalent to the solution of the Boltzmann equation, because having an infinite number of particles allows each particle to find a sample particle with precisely the same properties.

# Numerical experiment

To validate the proposed method, we conducted a numerical simulation of a relatively practical test case: a one-dimensional evaporation flow. A two-phase flow involving a phase change can be seen in several industrial applications, including desalination [25–29] and evaporative cooling [30–32]. These technologies utilize evaporation, which is the transfer of mass and heat from the liquid–gas interface, and an accurate evaluation of these transfers is essential for optimal design. These transfers are generated from a very thin layer near the liquid surface called the Knudsen layer, which has a thickness of several mean free paths [33]. As shown in Figure 2, we calculated the one-dimensional evaporative flow inside the Knudsen layer because the flow is usually low in velocity and near equilibrium, which matches the conditions for which the proposed method is intended. We assumed that the temperature of the liquid surface $T_L$ is a constant 80 °C and that the non-dimensionalized macroscopic flow velocity in the gas bulk $u_\infty^* = u_\infty/\sqrt{RT_L} = u_\infty/403.7$ is the input parameter. The ratio of inelastic collisions over all collisions $\phi$ of water molecules was set to 0.3. At the liquid surface, we imposed a completely diffusive boundary condition with both the evaporation and condensation coefficients set to unity. Additionally, we applied an outlet boundary condition to the far field. The full details of implementing these boundary conditions are discussed in our previous paper [21]. The selected discretization parameters of the problem were set to a cell width $\Delta x = 0.1\lambda = 1.24831 \times 10^{-8}$m and time step $\Delta t = 0.3\tau = 5.81292 \times 10^{-11}$s, where $\lambda$ is the mean free path of water and $\tau$ is the mean free time of water, both at the wall temperature $T_w$. For the deviational particle simulation, we set equilibrium distributions in the far field as the reference distributions. The temperature and molecular density at equilibrium were obtained via DSMC simulations, and the reference distributions were obtained by using equation (3).

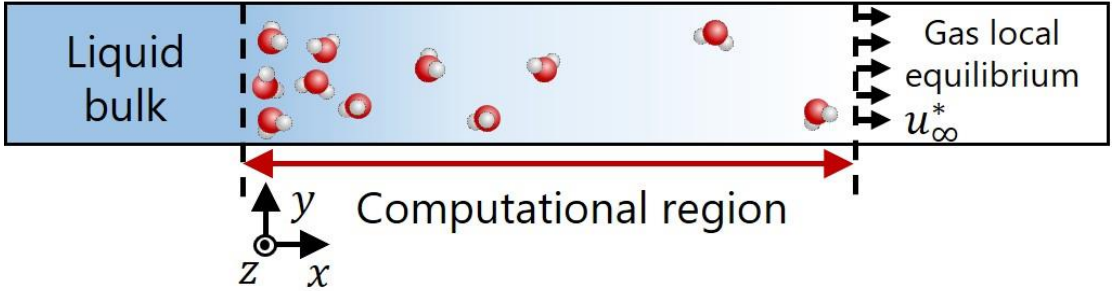

**Fig. 2** Schematic of the system in the numerical simulation. The temperature on the liquid surface was set to a constant ($T_w = 353.15$ K) as the boundary condition, and the non-dimensional velocity in the far field $u_\infty^*$ was set as a parameter.

We utilized this simulation setup to perform a parametric study and explore the sensitivity of the proposed method to different parameters for one test case. Then, we varied $u_\infty^*$ and conducted a degree of non-equilibrium study to assess the effectiveness of our approach across various flow scenarios. The temperature and molar density were calculated from the properties of sample particles. The molar density was obtained as

$$n = n_0 + \frac{g}{V_{\text{cell}}} \sum_{i=1}^{N_{\text{cell}}} \sigma_i, \tag{37}$$

where $V_{\text{cell}} = \Delta x$ is the volume of a computational cell. Two temperatures with different definitions were used. $T_{\text{rot}}$ is the rotational temperature that is calculated only from the rotational energy distribution:

$$T_{\text{rot}} = \frac{\int_0^\infty f(\varepsilon)d\varepsilon}{\xi \cdot n} = \frac{n_0 T_0 + \frac{2}{\xi k_b} \cdot \frac{g}{V_{cell}} \cdot \sum_{i=1}^{N_{\text{cell}}} \sigma_i \varepsilon_i}{n}, \tag{38}$$

$T$ is the temperature defined by both the translational and rotational energy, which is often simply called the temperature:

$$T = \frac{3T_{\text{tr}} + \xi T_{\text{rot}}}{3 + \xi}. \tag{39}$$

where $T_{\text{tr}}$ is the kinetic temperature and is defined as

$$T_{\text{tr}} = \frac{m}{3nk_b}\langle|\mathbf{c}-\bar{\mathbf{c}}|^2\rangle = \frac{n_0 T_0 + \frac{m}{3k_b} \cdot \frac{1}{V_{cell}} \cdot g \sum_{i=1}^{N_{\text{cell}}} \sigma_i |\mathbf{c}_i - \bar{\mathbf{c}}|^2}{n}. \tag{40}$$

## Parametric study

The proposed method requires two parameters as simulation constants: the weight of a sample particle $g$ and the width of the approximated delta function $\delta\varepsilon$ introduced in equation (31). Similar to mesh studies in computational fluid dynamics, these parameters must be set meticulously to achieve accurate and efficient simulation results. With this motivation, we investigated the impact of these parameters on the convergence error and computational cost. Specifically, we analyzed the convergence error in the temperature and the number of generated sample particles at the liquid surface and in the far field as representative metrics of the one-dimensional evaporation flow with $u_\infty^* = 0.04$ using various pairs of these simulation parameters. A slow flow velocity was chosen to minimize the effects of linearization of the governing equation. Two values were used to quantify the convergence error: the error in the far field $T_{\text{error},\infty}^*$ and maximum error in the whole computational domain $T_{\text{error,max}}^*$. These were non-dimensionalized by the temperature difference along the flow as

$$T_{\text{error},\infty}^* = \frac{|T_{\text{DP},\infty} - T_{\text{DSMC},\infty}|}{T_{\text{w}} - T_{\text{DSMC},\infty}}, \tag{41}$$

$$T_{\text{error,max}}^* = \frac{\max(|T_{DP} - T_{DSMC}|)}{T_{\text{w}} - T_{\text{DSMC},\infty}}, \tag{42}$$

where $T_{\text{DP}}, T_{\text{DSMC}}$ are the temperatures calculated by the proposed method and the reference DSMC simulation, respectively. $T_{\text{error},\infty}^*$ was used to evaluate the accuracy for the overall transfer properties from the Knudsen layer, and $T_{\text{error,max}}^*$ was used to evaluate the capability of tracking the non-equilibrium behavior of the flow. For the $\delta\varepsilon$ analysis, only $\delta\varepsilon$ was varied while the value of $g$ was maintained suitably small. For the weight analysis, $\delta\varepsilon$ was kept at a sufficiently small constant value while $g$ was systematically varied. Figure 3 shows the number of sample particles in a cell at the liquid surface and in the far field as well as the convergence error with respect to $\delta\varepsilon$. As $\delta\varepsilon$ increased, particle cancelation with the approximated delta function occurred more frequently, which led to fewer particles within a cell. Notably, in the small $\delta\varepsilon$ region where $\delta\varepsilon < 5 \times 10^{-21} J$, this effect was particularly pronounced and led to an exponential decrease in $N_{\text{cell}}$.

Simultaneously, a coarser approximation of the delta function introduced larger errors in the rotational energy with increasing $\delta\varepsilon$. In the same small $\delta\varepsilon$ region, $T^*_{\text{error},\infty}$ converged to the DSMC results, and even the maximum error $T^*_{\text{error,max}}$ was approximately 0.01 (i.e., 1%). Therefore, there is a tradeoff between computational efficiency and accuracy when selecting the value of $\delta\varepsilon$. To strike a balance between these two factors, a value of $\delta\varepsilon = 3 \times 10^{-21} J$ appears suitable.

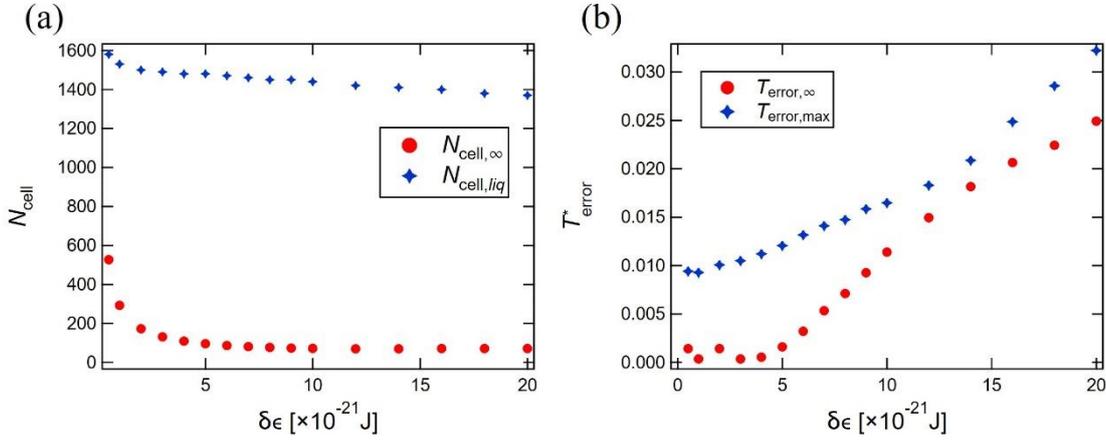

**Fig. 3** (a) Number of sample particles generated near the liquid surface (blue marker) and in the far field (red marker), and (b) non-dimensional convergence error of temperatures as functions of the simulation parameter $\delta\varepsilon$.

Figure 4 plots the same variables as functions of the weight of a sample particle. At the liquid surface, more particles are generated within a cell with a smaller weight $g$, and the relationship is nonlinear. This nonlinear behavior can be attributed to the inelastic collision process in the proposed method, which generates additional particles when the group reduction algorithm fails to conserve energy of the system. When $g$ is small and many particles are generated, collision events occur more frequently, which is further increased by the inelastic collisions. In the far field, the number of particles generated is independent of the parameter $g$. This value can be considered the minimum required number of sample particles in a cell for the proposed method. While theoretically no particles are necessary for calculating the far field, the minimum nonzero particle count arises from the limitation of particle cancelation in the proposed method. Specifically, the group reduction algorithm performs cancelation based solely on a snapshot of the particle distribution in the system without accounting for particles generated by collisions in the past or future, unlike the LVDSMC method and our elastic collision process. For the convergence error, we observed that the proposed method accurately computed transfer properties with $T^*_{\text{error},\infty}$ below 0.001 when $g$ was less than $5 \times 10^{12}$. In this region, non-equilibrium behavior was also captured faithfully with a maximum error $T^*_{\text{error,max}}$ of approximately 0.01. Again, a tradeoff exists, and $g = 2 \times 10^{12}$ seems suitable to ensure both computational efficiency and accuracy.

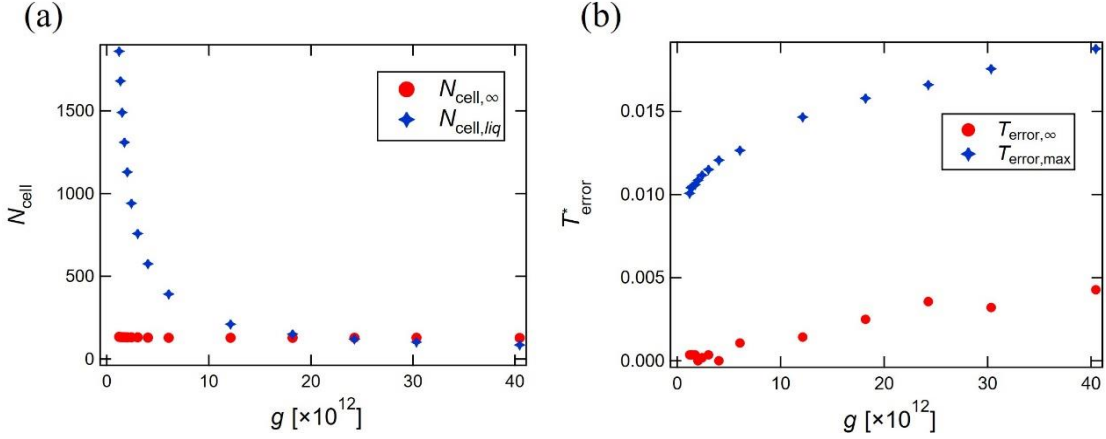

**Fig. 4** (a) Number of sample particles generated near the liquid surface (blue marker) and in the far field (red marker), and (b) non-dimensional convergence error of the temperature as functions of the weight of a sample particle $g$.

## Degree of non-equilibrium study

The above parametric study revealed that judiciously chosen values for the simulation parameters allows the proposed method to solve near equilibrium flows with an accuracy comparable to that of the traditional DSMC method. In other words, the introduction of the group reduction algorithm and approximation of the delta function are permissible when appropriate values for these parameters are selected. Furthermore, the linearization of the governing equation remains valid as long as deviations remain significantly smaller than the reference conditions. However, questions remain over the point at which this linearization breaks down in terms of accuracy and the efficiency of the proposed method compared to conventional approaches. To address these questions, we conducted a degree of non-equilibrium study on the accuracy and computational costs of the same one-dimensional evaporation flow simulation with various evaporative flow velocities. The degree of non-equilibrium is quantified as the non-dimensional rate of change in the molar density along a flow:

$$\Delta n^* = \frac{n_w - n_\infty}{n_w}. \tag{43}$$

Just as in the parametric study, the accuracy was represented by the convergence error of the temperature. To assess the computational efficiency, we calculated the relative computational cost (RCC) as the ratio of the time spent by the proposed method to that of the DSMC method to achieve equivalent smoothness in their results:

$$RCC = \frac{N_{\text{DP}} \cdot \tau_{\text{DP}}}{N_{\text{DSMC}} \cdot \tau_{\text{DSMC}}}. \tag{44}$$

where $N$ is the sample number used to obtain results with a standard error less than a specified threshold $SE_{\text{thres}}$ and $\tau$ is the computational time for one time step. When the simulations using the proposed DP method and the DSMC method yield the same standard error $SE_{\text{thres}}$, the following relation holds:

$$SE_{\text{thres}} = \frac{SD_{\text{DP}}}{\sqrt{N_{\text{DP}}}} = \frac{SD_{\text{DSMC}}}{\sqrt{N_{\text{DSMC}}}}. \tag{45}$$

where $SD$ is the standard deviation of the simulation results. With equation (45), $RCC$ can be rewritten as a function of the standard deviation and computational time for one time step:

$$RCC = \left(\frac{SD_{\text{DP}}}{SD_{\text{DSMC}}}\right)^2 \cdot \frac{\tau_{\text{DP}}}{\tau_{\text{DSMC}}}. \tag{46}$$

In this study, we used the standard error and standard deviation of the temperature in the far field to calculate the RCC. Figure 5 presents the results for the convergence error. Notably, the proposed method demonstrated an accurate simulation performance across a relatively wide range of $\Delta n^*$ values of up to approximately $\Delta n^* = 0.5$ for non-equilibrium behavior and up to $\Delta n^* = 0.6$ for the transfer properties. The difference between these two critical $\Delta n^*$ can be attributed to how we set the reference distributions. Because we used the equilibrium distribution in the far field as the reference, the deviations in the far field were relatively small. The linearized collision term solely considers interactions between deviations and the reference distribution. Thus, for such small deviations the reference distribution acts as a reservoir to absorb convergence errors in the far field, which results in smaller errors overall. However, as $\Delta n^*$ increases further and deviations near the liquid surface become large, errors due to linearization become more apparent.

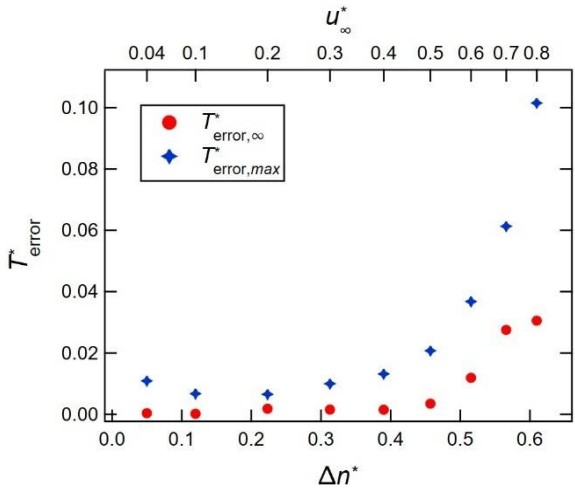

**Fig. 5** Non-dimensional convergence error of the temperature as a function of the degree of non-equilibrium ($\Delta n^*$).

As further validation of the proposed method, Figure 6 compares its simulation results for the evaporative velocity, temperature, and molar density along the flow at low ($u_\infty^* = 0.04$), middle ($u_\infty^* = 0.4$), and high ($u_\infty^* = 0.8$) velocities with those of the reference DSMC method. For the low-velocity flow, the degree of non-equilibrium $\Delta n^*$ was 0.05, which corresponds to the first point in Figure 5. With such a small deviation from the equilibrium distribution, linearization of the governing equation had little impact on the overall simulation results, and the proposed method accurately predicted the temperature and density distribution in all regions. Notably, the close agreement between the computed and reference values for both $T$ and $T_{\text{rot}}$ indicates that the energy exchange between different modes was accurately calculated, which supports the validity of our inelastic collision model and group reduction algorithm.

For the middle-velocity flow ($u_\infty^* = 0.4$), $\Delta n^*$ was 0.39, which is about one order of magnitude higher than for the low-velocity flow. Despite this relatively high $\Delta n^*$, the proposed method effectively captured trends in both the temperature and molar density. Figure 5 shows that the linearization with the assumption $f_d \ll f_0$ started to break down in this region of $\Delta n^*$ with $T_{\text{error,max}}^* = 0.02$. Nevertheless, the results obtained by the proposed method and DSMC method remained indistinguishable, which indicates that this level of error is acceptable for most practical scenarios.

For the high-velocity flow, $\Delta n^*$ was 0.61, which means that the molar density dropped by about 40% in the Knudsen layer. In such highly non-equilibrium flows, linearization is no longer valid, and the proposed method failed to accurately estimate the temperature. Especially near the liquid surface, relaxation was significantly underestimated because collisions between deviational particles, which now had a substantial impact on the distribution functions, were not calculated.

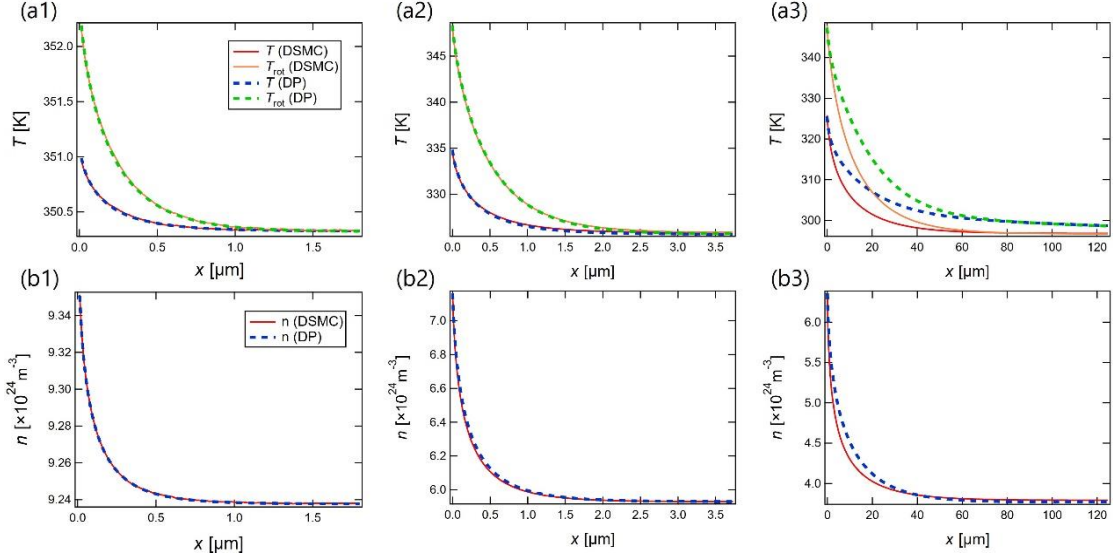

**Fig. 6** (a1, a2, a3) Temperature $T$ and rotational temperature $T_{\text{rot}}$, and (b1, b2, b3) molar density along the flow with the macroscopic flow velocities (a1, b1) $u_\infty^* = 0.04$, (a2, b2) $u_\infty^* = 0.4$, (a3, b3) $u_\infty^* = 0.8$. The solid lines show the reference results obtained by the DSMC method, and the dotted lines show the results obtained by the proposed method.

Figure 7 plots the RCC as a function of $\Delta n^*$. The RCC remained below unity up to approximately $\Delta n^* = 0.22$, which indicates that the proposed method is more efficient than the DSMC method within this range. Notably, at low $\Delta n^*$ of up to 0.1, the proposed method was 10–25 times faster than the DSMC method. However, the RCC increased rapidly with increasing $\Delta n^*$, and the proposed method became several times slower than the DSMC method at high $\Delta n^*$.

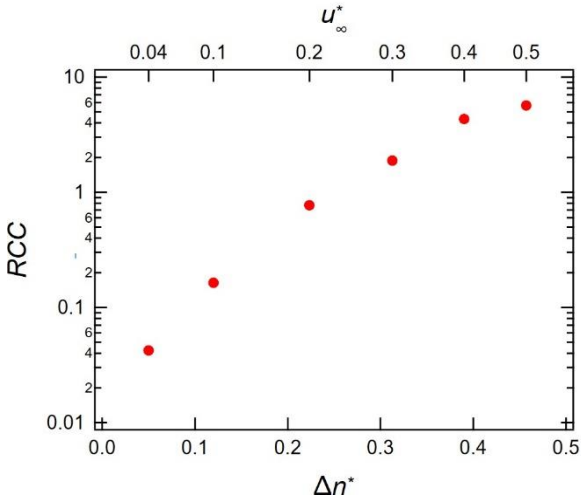

**Fig. 7** Relative computational cost as a function of the degree of non-equilibrium ($\Delta n^*$).

Table 1 presents two key values used to calculate the RCC: the standard deviation of the estimated temperature in the far field and the computational time for one time step. The data suggest that the proposed method significantly reduced the variance in all cases with approximately 5% of the variance observed with the DSMC method. According to the definition of the RCC in equation (46), this implies that calculations could be two to three orders of magnitude faster if both methods have the same computational time for one time step. However, the proposed method currently requires considerably more computational time for one time step than the DSMC method. Thus, it is only one order of magnitude faster at low $\Delta n^*$ and has a higher computational cost at high $\Delta n^*$. The extended computational time is primarily due to the group reduction algorithm in the inelastic collision operator. Specifically, selecting a particle from the computational cell is time-consuming for simulations of highly non-equilibrium flows because it involves searching through a larger number of particles generated by the greater degree of non-equilibrium. This search has an order of complexity $O(n\log n)$, which leads to a significant increase in the computational time for one time step ($\tau$) with increasing $\Delta n^*$. These results suggest that our strategy of combining the DP method with the LB model to reduce variance works effectively without significantly compromising accuracy. However, there is considerable potential for improving the implementation, particularly with regard to the inelastic collision operator. Future work will focus on enhancing the computational efficiency of the cancelation technique or even modifying the collision operator.

| $u_\infty^*$ | 0.04 | 0.1 | 0.2 | 0.3 | 0.4 | 0.5 |
|---|---|---|---|---|---|---|
| $\Delta n^*$ | 0.05 | 0.12 | 0.22 | 0.31 | 0.39 | 0.46 |
| $SD_{DP}$ | 0.0961 | 0.113 | 0.122 | 0.130 | 0.133 | 0.125 |
| $SD_{DSMC}$ | 2.07 | 2.12 | 2.10 | 2.32 | 2.40 | 2.49 |
| $\tau_{DP}$ | 0.142 | 0.629 | 3.60 | 10.6 | 21.9 | 42.1 |
| $\tau_{DSMC}$ | 0.00728 | 0.0109 | 0.0158 | 0.0180 | 0.0157 | 0.0203 |
| $RCC$ | 0.0422 | 0.163 | 0.770 | 1.88 | 4.31 | 5.66 |

**Table 1** Standard deviation ($SD$), computational time for one time step ($\tau$), and relative computational cost ($RCC$) in relation to the degree of non-equilibrium ($\Delta n^*$).

# Conclusion

Our proposed method for simulating polyatomic gas flows numerically solves the linearized governing equation for the DP method. The overall structure is based on the LVDSMC method, and the collision process is implemented as a combination of the source–sink process, LB model, and group reduction algorithm. To validate our approach, we carried out a numerical experiment and compared the results of the proposed method with those obtained by the reference DSMC method for a one-dimensional evaporation flow of water. The results showed that selecting appropriate values for two simulation parameters (i.e., the weight of sample particles $g$ and the width of the approximated delta function $\delta\varepsilon$ used to identify two rotational energies) allowed the proposed method to accurately estimate the macroscopic variables of the flow with higher computational efficiency than the DSMC method for a relatively wide range of the degree of non-equilibrium $\Delta n^*$. Crucially, the proposed method effectively reduces the statistical variance, which demonstrates the potential of combining the DP method and LB model to efficiently simulate polyatomic gas flows. While the proposed method incorporates a straightforward approach of adding a particle cancelation technique to the general collision operator, this comes with additional computational costs. Future work will involve formulating a source–sink process that is applicable to the inelastic collision process to further improve the computational efficiency.

# References

[1]   P. Degond and B. Lucquin-Desreux, *The Fokker-Planck Asymptotics of The Boltzmann Collision Operator In The Coulomb Case*, Math. Model. Methods Appl. Sci. **02**, 167 (1992).

[2]   E. P. Gross and E. A. Jackson, *Kinetic Models and the Linearized Boltzmann Equation*, Phys. Fluids **2**, 432 (1959).

[3]   C. D. Boley and S. Yip, *Modeling Theory of the Linearized Collision Operator for a Gas Mixture*, Phys. Fluids **15**, 1424 (1972).

[4]   Z. Cai, Y. Fan, and R. Li, *A Framework on Moment Model Reduction for Kinetic Equation*, SIAM J. Appl. Math. **75**, 2001 (2015).

[5]   M. Torrilhon, *Modeling Nonequilibrium Gas Flow Based on Moment Equations*, Annu. Rev. Fluid Mech. **48**, 429 (2016).

[6]   P. L. Bhatnagar, E. P. Gross, and M. Krook, *A Model for Collision Processes in Gases. I. Small Amplitude Processes in Charged and Neutral One-Component Systems*, Phys. Rev. **94**, 511 (1954).

[7]   G. A. Bird, *Molecular Gas Dynamics and the Direct Simulation of Gas Flows* (Oxford university press, 1994).

[8]   J. E. Broadwell, *Study of Rarefied Shear Flow by the Discrete Velocity Method*, J. Fluid Mech. **19**, 401 (1964).

[9]   F. G. Tcheremissine, *Conservative Evaluation of Boltzmann Collision Integral in Discrete Ordinates Approximation*, Comput. Math. with Appl. **35**, 215 (1998).

[10]  T. Ohwada, Y. Sone, and K. Aoki, *Numerical Analysis of the Shear and Thermal Creep Flows of a Rarefied Gas over a Plane Wall on the Basis of the Linearized Boltzmann Equation for Hard-sphere Molecules*, Phys. Fluids A Fluid Dyn. **1**, 1588 (1998).

[11]  A. Rana, M. Torrilhon, and H. Struchtrup, *A Robust Numerical Method for the R13 Equations of Rarefied Gas Dynamics: Application to Lid Driven Cavity*, J. Comput. Phys. **236**, 169 (2013).

[12]  T. M. M. Homolle and N. G. Hadjiconstantinou, *Low-Variance Deviational Simulation Monte Carlo*, Phys.

Fluids **19**, 041701 (2007).

[13]  T. M. M. Homolle and N. G. Hadjiconstantinou, *A Low-Variance Deviational Simulation Monte Carlo for the Boltzmann Equation*, J. Comput. Phys. **226**, 2341 (2007).

[14]  G. A. Radtke and N. G. Hadjiconstantinou, *Variance-Reduced Particle Simulation of the Boltzmann Transport Equation in the Relaxation-Time Approximation*, Phys. Rev. E - Stat. Nonlinear, Soft Matter Phys. **79**, 056711 (2009).

[15]  W. Wagner, *Deviational Particle Monte Carlo for the Boltzmann Equation*, Monte Carlo Methods Appl. **14**, 191 (2008).

[16]  H. A. Al-Mohssen and N. G. Hadjiconstantinou, *Low-Variance Direct Monte Carlo Simulations Using Importance Weights*, ESAIM Math. Model. Numer. Anal. **44**, 1069 (2010).

[17]  H. A. Al-Mohssen and N. G. Hadjiconstantinou, *A Practical Variance Reduced DSMC Method*, AIP Conf. Proc. **1333**, 219 (2011).

[18]  M. T. Casella, George, Robert, Christian P., *Generalized Accept-Reject Sampling Schemes*, Inst. Math. Stat. Lect. Notes-Monograph Ser. **45**, 342 (2004).

[19]  R. M. Neal, *Slice Sampling*, Ann. Stat. **31**, 705 (2003).

[20]  G. E. P. Box and M. E. Muller, *A Note on the Generation of Random Normal Deviates*, Ann. Math. Stat. **29**, 610 (1958).

[21]  H. Matsumoto, T. Hori, Y. Yoshimoto, and I. Kinefuchi, *Evaporation Knudsen Layer Analysis by the Low-Variance Deviational Simulation Monte Carlo Method*, in prepration.

[22]  C. Borgnakke and P. S. Larsen, *Statistical Collision Model for Monte Carlo Simulation of Polyatomic Gas Mixture*, J. Comput. Phys. **18**, 405 (1975).

[23]  I. D. Boyd, *Temperature Dependence of Rotational Relaxation in Shock Waves of Nitrogen*, J. Fluid Mech. **246**, 343 (1993).

[24]  I. D. Boyd, *Analysis of Rotational Nonequilibrium in Standing Shock Waves of Nitrogen*, AIAA J. **28**, 1997 (1990).

[25]  J. E. MILLER, Review of Water Resources and Desalination Technologies, 2003.

[26]  T. Humplik et al., *Nanostructured Materials for Water Desalination*, Nanotechnology **22**, (2011).

[27]  A. Deshmukh, C. Boo, V. Karanikola, S. Lin, A. P. Straub, T. Tong, D. M. Warsinger, and M. Elimelech, *Membrane Distillation at the Water-Energy Nexus: Limits, Opportunities, and Challenges*, Energy Environ. Sci. **11**, 1177 (2018).

[28]  W. Chen, S. Chen, T. Liang, Q. Zhang, Z. Fan, H. Yin, K. W. Huang, X. Zhang, Z. Lai, and P. Sheng, *High-Flux*

*Water Desalination with Interfacial Salt Sieving Effect in Nanoporous Carbon Composite Membranes*, Nat. Nanotechnol. (2018).

[29] J. Lee and R. Karnik, *Desalination of Water by Vapor-Phase Transport through Hydrophobic Nanopores*, in *Journal of Applied Physics* (2010).

[30] M. Bongarala, H. Hu, J. A. Weibel, and S. V Garimella, *Microlayer Evaporation Governs Heat Transfer Enhancement during Pool Boiling from Microstructured Surfaces Microlayer Evaporation Governs Heat Transfer Enhancement during Pool Boiling from Microstructured Surfaces*, Appl. Phys. Lett. **221602**, (2022).

[31] Z. Lu, T. R. Salamon, S. Narayanan, K. R. Bagnall, D. F. Hanks, D. S. Antao, B. Barabadi, J. Sircar, M. E. Simon, and E. N. Wang, *Design and Modeling of Membrane-Based Evaporative Cooling Devices for Thermal Management of High Heat Fluxes*, IEEE Trans. Components, Packag. Manuf. Technol. **6**, 1056 (2016).

[32] Z. Lu, S. Narayanan, and E. N. Wang, *Modeling of Evaporation from Nanopores with Nonequilibrium and Nonlocal Effects*, Langmuir **31**, 9817 (2015).

[33] M. Knudsen, *Die Gesetze Der Molekularströmung Und Der Inneren Reibungsströmung Der Gase Durch Röhren*, Ann. Phys. **333**, 75 (1909).